\documentclass[%
 reprint,
 amsmath,amssymb,
 aps,
]{revtex4-1}

\usepackage{graphicx}
\usepackage{dcolumn}
\usepackage{bm}

\begin{document}


\title{Covariant Hamiltonian approach for time-dependent potentials applied to a pill-box cavity}

\author{E. Laface}
 \email{emanuele.laface@ess.eu}
\author{B.T. Folsom}%

\affiliation{%
 European Spallation Source ERIC\\
 Lund, Sweden
 }%


\date{\today}

\begin{abstract}
The common treatment of time-dependent potentials, such as those used for radio frequency cavities, is to average a potential's time component through the interval that the reference particle spends in the cavity. Such an approach, using the so-called transit-time factor, uses time as the independent variable in the Hamiltonian. In this paper, we instead propose a fully covariant Hamiltonian to treat the time component of the potential like any other space component. We show how to calculate the dynamics of the particles in a pill-box cavity using an explicit sympletic integrator. Finally, we compare the results with the simulator TraceWin.
\end{abstract}

\maketitle


\section{\label{sec:level1}Introduction}
When a charged particle passes through an accelerator element which generates a time-dependent field, this yields a Hamiltonian which is no longer a constant of motion. This requires special techniques to handle the time component, such as the transit-time factor (TTF) for RF cavities, where the time component is averaged and removed from the Hamiltonian.

The time dependency of the Hamiltonian can also be removed if we consider the covariant Hamiltonian, in a 8-dimensional phase space, where together with the usual 6-dimensions ($x,P_x$; $y,P_y$; $z,P_z$) we also have time and its canonical conjugate $(t, P_t)$. In this phase space, the Hamiltonan $H=H(x,y,z,t,P_x,P_y,P_z,P_t)$ is always a constant of motion and will satisfy four pairs of Hamilton equations
\begin{eqnarray}
    \frac{dx^\mu}{d\tau}&=&\frac{\partial H}{\partial P^\mu}\\
    \frac{dP^\mu}{d\tau}&=-&\frac{\partial H}{\partial x^\mu}
\end{eqnarray}
where $\tau$ is the``proper'' time of the particle, that is the time in a reference frame that moves with the particle; $x^\mu=ct,x,y,z$ when $\mu=0,1,2,3$; and $P^\mu$, defined rigorously in the next section, are the corresponding momenta.

The solutions to these Hamilton's equations can be obtained in many ways, the technique used to solve them depends on the form of the potential used. Expanding on preliminary studies~\cite{folsom-thesis}, the potential considered in the example of this paper is
a simple pill-box cavity. Such a potential can be split as explained in \cite{wu2003}, so we will adopt that technique to have an explicit symplectic integrator.

The results obtained with this method will be compared with the standard simulator used at the European Spallation Source, TraceWin \cite{uriot2011tracewin}.

\section{Hamiltonian}
The Lorentz invariant
\begin{equation}
    \label{invariant}
    ds^2=-c^2dt^2+dx^2+dy^2+dz^2=c^2d\tau^2
\end{equation}
leads to the Lagrangian
\begin{equation}
    L=\frac{m}{2}v_\mu v^\mu
\end{equation}
where
\begin{eqnarray}
    v^\mu&=&\left(c\frac{dt}{d\tau}, \frac{dx}{d\tau}, \frac{dy}{d\tau}, \frac{dz}{d\tau}\right)\\
    v_\mu&=&\left(-c\frac{dt}{d\tau}, \frac{dx}{d\tau}, \frac{dy}{d\tau}, \frac{dz}{d\tau}\right).
\end{eqnarray}
With an external electromagnetic quadri-potential $A^\mu=\left(\frac{\phi}{c},A_x,A_y,A_z\right)$ the Lagrangian is \cite{Landau1980Classical}
\begin{equation}
    L=\frac{m}{2}v_\mu v^\mu+\epsilon v^\mu A_\mu
\end{equation}
where $\epsilon$ is the electric charge of the particle. The corresponding Hamiltonian is calculated with the Legendre transform as
\begin{eqnarray}
\label{cov_momenta}
    P_\mu &=& \frac{\partial L}{\partial v^\mu}\\
    H&=&P_\mu v^\mu-L.
\end{eqnarray}
This Hamiltonian generates eight equations of motion with the constraint $v_\mu v^\mu=c^2$ so not all the variables are independent.

\section{The Pill-Box cavity}
The case discussed here is the simple pill-box cavity with the simplest accelerating mode (TM$_{01}$). The electric and magnetic fields, expressed in cylindrical coordinates plus time $(t, r, \theta, z)$ are \cite{andrzej2014beam}
\begin{eqnarray}
    E_z=E_0J_0\left(\frac{b_{01}}{a}r\right)\sin\left(\omega t+\phi_0\right)\\
    B_\theta=\frac{E_0}{c}J_1\left(\frac{b_{01}}{a}r\right)\cos(\omega t+\phi_0)\\
    E_r=E_\theta=B_r=B_z=0.
\end{eqnarray}
$J_0$ and $J_1$ are the Bessel functions of kind $0$ and $1$; $b_{01}$ is the first zero of $J_0$, that is, $J_0(b_{01})=0$; $a$ is the radius of the pill box cavity (this leaves the electric field as zero when $r=a$); $\omega$ is the oscillation frequency of the field in the cavity multiplied by $2\pi$ which is fixed by the cavity aperture through the relationship $\omega=c\frac{b_{01}}{a}$; and $\phi_0$ is the phase that the particle sees when it arrives to the entrance of the cavity.

These fields can be expressed as a potential:
\begin{eqnarray}
    A_r=A_\theta=0\\
    A_z=\frac{E_0}{\omega}J_0\left(\frac{b_{01}}{a}r\right)\cos(\omega t+\phi_0).
\end{eqnarray}

Because the potential is only in the $z$ direction, we do not need to transform the Lagrangian in cylindrical coordinates, but we can use the potential directly in the calculations recalling that $r=\sqrt{x^2+y^2}$.
The Lagrangian is
\begin{eqnarray}
    L=\frac{m}{2}\left(v_0v^0+v_1v^1+v_2v^2+v_3v^3\right)+\epsilon v^3A_3.
\end{eqnarray}

The conjugate momenta are then
\begin{eqnarray}
    P_0 &=& \frac{\partial L}{\partial v^0}=mv_0\\
    P_1 &=& \frac{\partial L}{\partial v^1}=mv_1\\
    P_2 &=& \frac{\partial L}{\partial v^2}=mv_2\\
    P_3 &=& \frac{\partial L}{\partial v^3}=mv_3+\epsilon A_3
\end{eqnarray}
so we can rewrite the 4-velocity as
\begin{eqnarray}
    v^\mu&=&\left(\frac{P^0}{m}, \frac{P^1}{m},\frac{P^2}{m},\frac{P^3-\epsilon A^3}{m} \right).
\end{eqnarray}
The Hamiltonian is
\begin{eqnarray}
    H&=&\frac{P_0P^0}{m}+\frac{P_1P^1}{m}+\frac{P_2P^2}{m}+\frac{P_3(P^3-\epsilon A^3)}{m}-L\\
    \nonumber
    &=&\frac{P_0P^0}{2m}+\frac{P_1P^1}{2m}+\frac{P_2P^2}{2m}+\frac{(P_3-\epsilon A_3)(P^3-\epsilon A^3)}{2m}
\end{eqnarray}
or explicitly
\begin{eqnarray}
    \label{hamiltonian}
    H&=&-\frac{P^2_t}{2m}+\frac{P^2_x}{2m}+\frac{P^2_y}{2m}+\frac{(P_z-\epsilon A_z)^2}{2m}.
\end{eqnarray}

The Hamiltonian (\ref{hamiltonian}) can be treated with the explicit symplectic integrator first developed in \cite{wu2003} and discussed in \cite{Dragt2019} Chapter 12 Section 9.

\section{The algorithm}
The only problematic term in the Hamiltonian (\ref{hamiltonian}) 
is the one dependent on $z$ because it mixes the momentum $P_z$ with the position contained in $A_z$. The idea of our integrator is to create a new function $U_z(t,x,y,z)$ such that $A_z=\frac{\partial U_z}{\partial z}$. In our case
\begin{equation}
    U_z=\frac{E_0}{\omega}J_0\left(\frac{b_{01}}{a}r\right)\cos(\omega t+\phi_0)z.
\end{equation}

The Lie transform of $U_z$ has the property:
\begin{eqnarray}
e^{-\epsilon :U_z:}z&=&z\\
e^{-\epsilon :U_z:}P_z&=&P_z-\epsilon \frac{\partial U_z}{\partial z}
\end{eqnarray}
and as consequence we have
\begin{equation}
e^{-\epsilon :U_z:}
e^{-\frac{h}{2}:\frac{P^2_z}{2m}:}
e^{\epsilon :U_z:}=e^{-\frac{h}{2}:\frac{(P_z-\epsilon A_z)^2}{2m}:}.
\label{gauge}
\end{equation}
This splitting technique is similar to the usual drift-kick-drift, but here the term $U_z$ can be seen as a gauge transformation instead of a kick. Calling $K_t=-\frac{P^2_t}{2m}$ and $K_i=\frac{P^2_i}{2m}$ with $i=x,y,z$ we have the second order explicit symplectic integrator as

\begin{eqnarray}
\label{algo}
\nonumber
S_2 &=& e^{-\frac{h}{2}:K_x:}
e^{-\frac{h}{2}:K_y:}
e^{-\epsilon :U_z:}
e^{-\frac{h}{2}:K_z:}
e^{\epsilon :U_z:}\times\\
\nonumber
&&e^{-h:K_t:}\times\\
&&e^{-\epsilon :U_z:}
e^{-\frac{h}{2}:K_z:}
e^{\epsilon :U_z:}
e^{-\frac{h}{2}:K_y:}
e^{-\frac{h}{2}:K_x:}.
\end{eqnarray}
This integrator can be extended to higher order integrators applying the technique of Yoshida \cite{Yoshida:1990zz} or that of Suzuki \cite{SUZUKI1990319}. Both techniques are explored in detail in \cite{Hairer:1250576}.

Every step of the integrator has to be evaluated on coordinates and momenta. The only terms different from the identity map are:
\begin{eqnarray}
\label{integrator}
\nonumber
e^{-\frac{h}{2}:K_i:}i&=&i+\frac{h}{2m}P_i~{\rm{for}}~i=x,y,z\\
\nonumber
e^{-h:K_t:}t&=&t-\frac{h}{m}P_t\\
\nonumber
e^{\epsilon :U_z:}P_x&=&P_x-\frac{\epsilon xzE_0b_{01}}{ar\omega}J_1\left(\frac{b_{01}}{a}r\right)\cos(\omega t+\phi_0)\\
\nonumber
e^{\epsilon :U_z:}P_y&=&P_y-\frac{\epsilon yzE_0b_{01}}{ar\omega}J_1\left(\frac{b_{01}}{a}r\right)\cos(\omega t+\phi_0)\\
\nonumber
e^{\epsilon :U_z:}P_z&=&P_z+\frac{\epsilon E_0}{\omega}J_0\left(\frac{b_{01}}{a}r\right)\cos(\omega t+\phi_0)\\
e^{\epsilon :U_z:}P_t&=&P_t-\epsilon zE_0J_0\left(\frac{b_{01}}{a}r\right)\sin(\omega t+\phi_0).
\end{eqnarray}

\section{Numerical results}
We compared the results of this symplectic integrator with the well-established code TraceWin~\cite{uriot2011tracewin}. In particular, we benchmarked against a drift-gap-drift model, where we used the ``bunched cavity or thin gap'' element of TraceWin for the gap. We also compared with a fieldmap model, but the discrepancies with a thin gap are negligible and not presented here.

In order to compare the results of this symplectic integrator with TraceWin, we need the results in the same reference frame. The first step is thus to find the transformation from the covariant Hamiltonian to the Hamiltonian in the laboratory frame that uses $t$ as the independent variable. This can be done using Eq.~(\ref{invariant}) noting that $\frac{dt}{d\tau}=\gamma$. From the definition of the momenta Eq.~(\ref{cov_momenta}) we have for the general Hamiltonian
\begin{eqnarray}
    P^0&=&mv^0+\epsilon A^0=mc\frac{dt}{d\tau}+\epsilon \frac{\phi}{c}\\
    P^i&=&mv^i+\epsilon A^i=m\frac{dx^i}{d\tau}+\epsilon A^i;~(i=1,2,3)
\end{eqnarray}
and this is
\begin{eqnarray}
    cP^0&=&\gamma mc^2+\epsilon \phi\\
    c^2\left(\vec{P}-\epsilon \vec{A}\right)^2&=&\gamma^2 m^2c^2v^2
\end{eqnarray}
where the vector notation refers to the three spatial coordinates and the square is the norm square of the vector. Using the fact that $v^2=\left(1-\frac{1}{\gamma^2}\right)$, we can connect the spatial component with the time component
\begin{eqnarray}
    \label{ham-time}
    cP^0=c\sqrt{\left(\vec{P}-\epsilon \vec{A}\right)^2+m^2c^2}+\epsilon \phi
\end{eqnarray}
this is the Hamiltonian when $t$ is used as an independent variable \cite{Landau1980Classical} if we identify $cP^0$ as the Hamiltonian.

From this Hamiltonian we can apply the transformations as in \cite{andrzej2014beam}. The first transformation is to normalize the left and right sides of Eq. (\ref{ham-time}) by the reference momentum $P_r=\beta \gamma mc$. This momentum is well established when the Hamiltonian does not depend on time, but it is not uniquely defined along the motion of a time-dependent Hamiltonian because $\beta$ and $\gamma$ are not constant. Thus, for every instance in which we compare a step of the covariant Hamiltonian algorithm with TraceWin, we need to normalize the result to the reference momentum of that step calculated from the total energy $cP^0$.

\begin{figure}
  \includegraphics[width=0.48\textwidth]{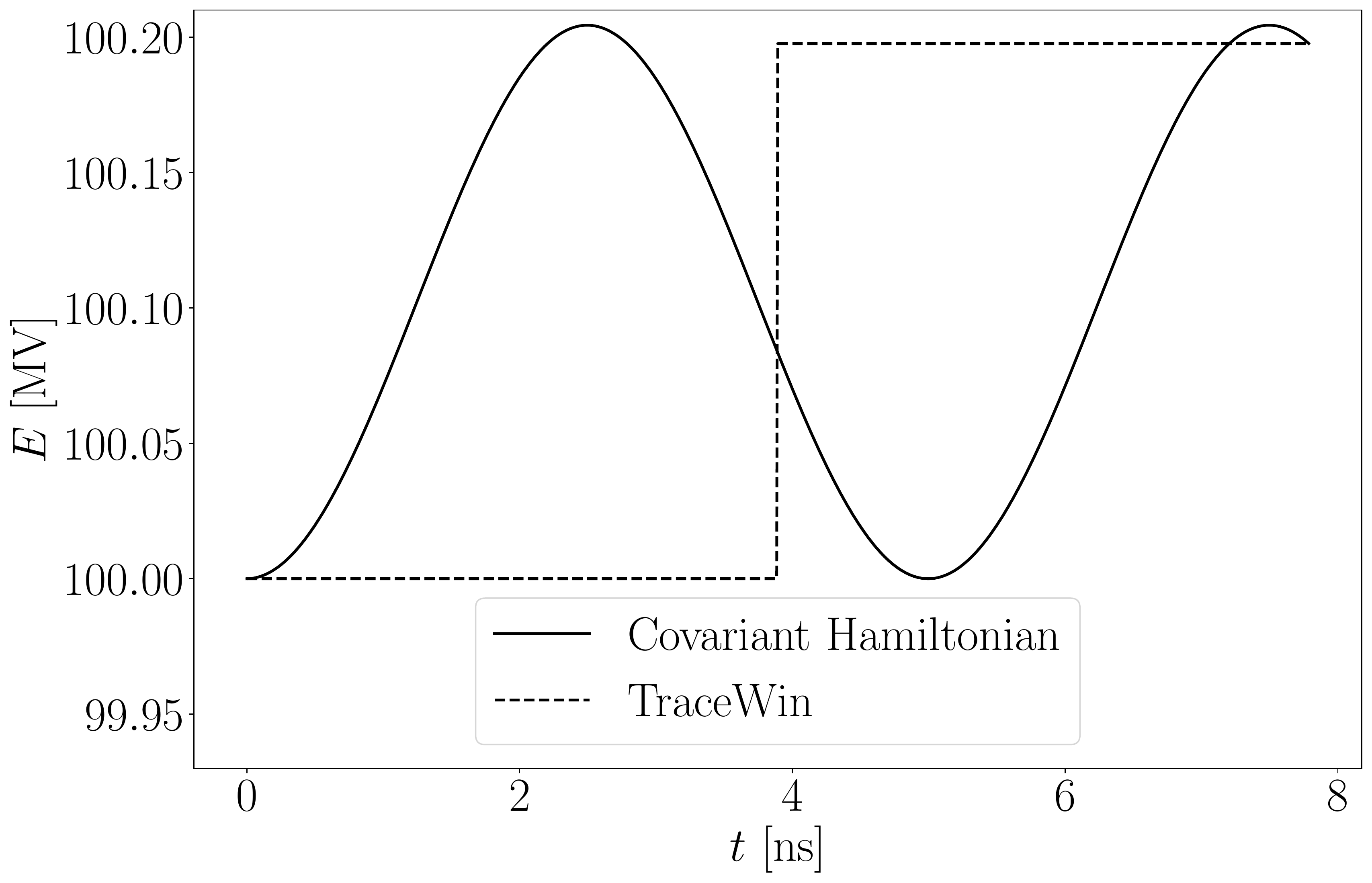}
  \caption{Energy of a particle on the z-axes with a kinetic energy of 100 MeV at the cavity entrance. Comparison between the covariant Hamiltonian algorithm and the step function in a classical TTF algorithm (such as the one used in TraceWin).}
  \label{energy}
\end{figure}

It is worth stressing here that the algorithm used for the integration of the covariant Hamiltonian is the one in Eqs. (\ref{algo}) and (\ref{integrator}) where no normalization is applied, and only the final result (or an intermediate result as with energy plot in Fig. \ref{energy}) is normalized for comparison with TraceWin. To introduce a normalization on each step may be useful to keep the numerical quantities small in the execution of the algorithm, but it introduces an additional dependence of the steps from the energy that does not come from the Hamiltonian itself and violates the symplectic nature of the Lie transform. This is why we use the normalization only on the final result for comparison purposes.

The last input we need for our simulator is the step size $h$. At the entrance of the cavity we have $\gamma_i$ which can be calculated from the initial kinetic energy as $\gamma_i=\frac{{E_k}_i}{mc^2}+1$ while the speed in the laboratory frame is $\sqrt{1-\frac{1}{\gamma^2_i}}c=\beta_ic$. Without acceleration, the time to traverse the first length step $dL$ of the pill-box cavity having a total length $L$ is $dL=\beta_icdt$ and we can again use the transformation $\frac{dt}{d\tau}=\gamma$ leaving
\begin{equation}
    dL=\beta_i\gamma_icd\tau.
\end{equation}
We can thus divide the length $L$ into segments and calculate the proper $h$ in Eqs. (\ref{integrator}) to evaluate the integrator. The $\beta_i$ and $\gamma_i$ must be recalculated at each step.

This manner of selecting $h$ may seem to introduce a dependence on the energy that can break the symplectic nature of the algorithm because it does originate from the Hamiltonian. This is not true for the integrator used here because the step length is pre-calculated and used as a constant for each step of pure drifts (the only part of Eqs. (\ref{integrator}) where $h$ appears) and we know that the Lie transform always generates a canonical (and so symplectic) map, regardless the size of the integration step.

The symplecticity is preserved by this procedure where the momentum is fixed along the drift sections, but another issue is a more subtle. The separation methods for symplectic integration, including the one used in this paper, minimize the error compared to the full Hamiltonian when the choice of the coefficients cancels the high-order terms of the Zassenhaus formula (which provides the expansion of the exponential of the Lie operator, see \cite{andrzej2014beam} p. 314). For our algorithm, this cancellation occurs when $h$ is unchanged from the first to the last drift of the $S_2$ integrator.

However, in our case we want to change the value of $h$ every time that we pass through a kick given by the $U_z$ potential. This procedure may break the symmetry of the $S_2$ integrator, which is no longer of second order, and introduces some error. To avoid this problem, one can use the same value for $h$ along one full step of integrating $S_2$, and then only recalculate $h$ between integration steps.

\begin{figure*}
  \includegraphics[width=\textwidth]{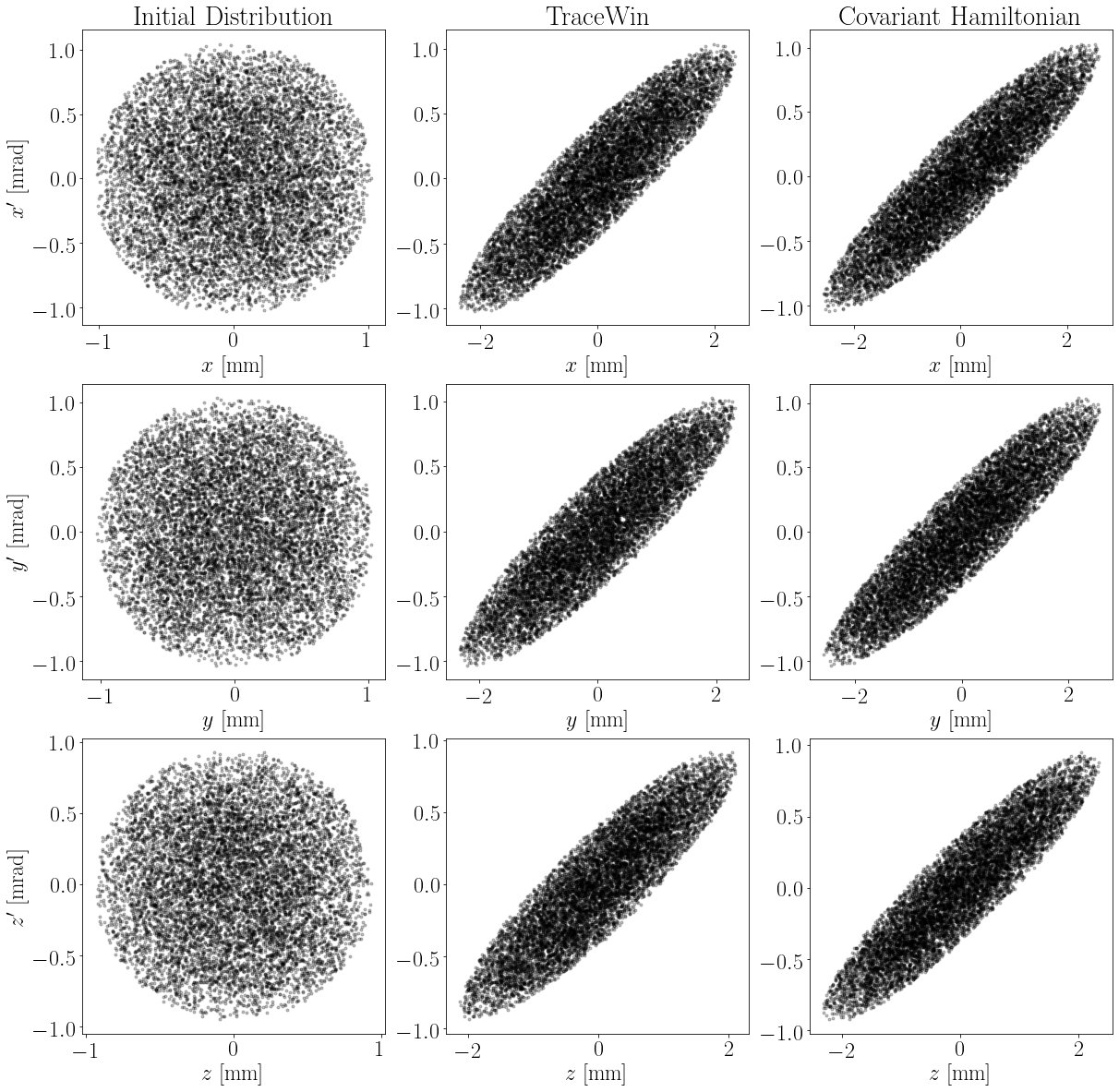}
  \caption{Comparison of the phase space on the three axes starting from the same initial distribution for TraceWin and our Covariant Hamiltonian algorithm.}
  \label{phase-space}
\end{figure*}

The test case we use for comparison with Tracewin is a bunch consisting of $10^5$ protons uniformly distributed around a kinetic energy of $100$~MeV and passing through a pill-box cavity with a frequency of $200$~MHz and a length of $1$~m. The phase space for the three spatial dimensions $x,y,z$ at the entrance and at the exit of the cavity is shown in Fig. \ref{phase-space}, where $z$ is collinear with the beamline. The comparison shows a strong matching, but this is not a surprise because such a cavity is primarily a drift space for our beam.

The interesting part here is in the behaviour of the energy. The parameters selected for this numerical test are such that the cavity is working quite far from any ideal acceleration conditions. In fact, a proton of $100$~MeV has a relativistic $\beta=0.428$ and will take $7.79$~ns to pass through the 1~m of length of the cavity. A cavity oscillating at $200$~MHz will do $1.5$ full oscillations in $7.79$~ns. This oscillation becomes traceable (Fig. \ref{energy}) in the energy gain of the covariant Hamiltonian algorithm if we track a reference particle with initial conditions set to zero for each coordinate and momenta component and with a kinetic energy of 100 MeV.

The transit-time factor of such a field is $~0.197$, that is, the acceleration is only 20\% of the peak field. In TraceWin, the effect will be to have a total energy increase of $~0.2$ MeV because the GAP element of this program uses the TTF to calculate the energy gain. In the simulator constructed with the covariant Hamiltonian, we will see the same increase in energy at the end, but we can see the integration along the length of the cavity for each integration step.

\section{conclusions}
We presented an algorithm for performing symplectic integration of Hamiltonian when the external potential depends on time. The main idea is to use the full covariant Hamiltonian and to integrate the field in 8-dimensional phase space applying a Lie-operator method to the field component as was done for s-dependent fields in \cite{wu2003}. A comparison with TraceWin, a well-established particle simulator, shows that the method reproduces the correct dynamics for the simple case of a pill-box cavity. More complicated potentials are feasible if they respect the separability expressed by Eq.~(\ref{gauge}).


\nocite{*}
\bibliography{biblio}

\end{document}